\documentclass[5p,11pt]{elsarticle}
\usepackage{amsfonts,color}
\usepackage{amsmath}
\usepackage{amssymb}
\usepackage{graphicx,dsfont}
\journal{Physics Letter B}

\newcommand{\be}{\begin{eqnarray}}
\newcommand{\ee}{\end{eqnarray}}

\newcommand{\bi}{\begin{itemize}}
\newcommand{\ei}{\end{itemize}}
\newcommand{\bea}{\begin {eqnarray}}
\newcommand{\eea}{\end{eqnarray}}

\def \to {\rightarrow}

\begin{document}
\begin{frontmatter}
\title{Pion-nucleus induced Drell-Yan cross section in models inspired by light-front holography}


\author[1,2,3,4]{Jiangshan Lan}
\ead{jiangshanlan@impcas.ac.cn}

\author[1,2,3]{Chandan Mondal\corref{cor1}}
\ead{mondal@impcas.ac.cn}

\address[1]{Institute of Modern Physics, Chinese Academy of Sciences, Lanzhou 730000, China}
\address[2]{CAS Key Laboratory of High Precision Nuclear Spectroscopy, Institute of Modern Physics, Chinese Academy of Sciences, Lanzhou 730000, China}
\address[3]{School of Nuclear Science and Technology, University of Chinese Academy of Sciences, Beijing 100049, China}
\address[4]{Lanzhou University, Lanzhou 730000, China}

\cortext[cor1]{Corresponding author}

\begin{abstract}
We evaluate the cross section for the pion induced Drell-Yan process in two different models of the pion light-front wave function, inspired by light-front holographic approaches. 
We obtain the valence quark distribution functions in those models, which after QCD evolution, are consistent with the experimental data from the E-0615 experiment at Fermilab. Supplemented by the parton distribution functions of the target nuclei, we obtain the cross section consistent with the experimental data for the $\pi^-$ Nucleus $\rightarrow {\mu^+\mu^-X}$ Drell-Yan process.
\end{abstract}


\end{frontmatter}

\section{Introduction} \label{intro}
The determination of parton distribution functions (PDFs) from the analysis of hard scattering processes has emerged as one of the main topics of hadron physics, attracting dedicated  theoretical and experimental efforts~\cite{Bordalo:1987cs,Freudenreich:1990mu,Sutton:1991ay, Conway:1989fs,Badier:1983mj,Detmold:2003tm,Gluck:1999xe,Wijesooriya:2005ir,Aicher:2010cb,Watanabe:2017pvl,Hecht:2000xa,Nam:2012vm,Holt:2010vj,Pumplin:2002vw,Ball:2017nwa,Alekhin:2017kpj,Dulat:2015mca,Harland-Lang:2014zoa,Izubuchi:2019lyk,Bednar:2018mtf,Barry:2018ort}. Quantum chromodynamics (QCD) is the underlying theory to explain the structure of hadrons in all energy regions. In the low energy region, quarks are confined and the dynamical breaking of the chiral symmetry emerges. This leads to pions having a small
mass when compared to other hadrons, playing the role of
the pseudoscalar Goldstone bosons.

The Drell-Yan dilepton production in $\pi^-$-tungsten reactions is one of the available experimental sources with access to the pion PDFs~\cite{Bordalo:1987cs,Freudenreich:1990mu,Sutton:1991ay}. The next-to-leading order (NLO) analyses of this Drell-Yan process have been carried out in Refs.~\cite{Sutton:1991ay,Gluck:1999xe,Wijesooriya:2005ir}, whereas the determinations of the light meson and the nucleon PDFs with associated uncertainties from the experiments have been reported in Refs.~\cite{Detmold:2003tm,Aicher:2010cb,Watanabe:2017pvl,Hecht:2000xa,Nam:2012vm,Holt:2010vj,Pumplin:2002vw,Ball:2017nwa,Alekhin:2017kpj,Dulat:2015mca,Harland-Lang:2014zoa}. The first global QCD analysis of the pion PDF has been performed in Ref.~\cite{Barry:2018ort}. The pion PDF has also been the subject of detailed analyses in lattice QCD~\cite{Detmold:2003tm,Martinelli:1987bh,Oehm:2018jvm,Abdel-Rehim:2015owa,Lin:2017snn,Brommel:2006zz,Sufian:2019bol}. It has also been investigated in phenomenological models such as the Nambu--Jona-Lasinio (NJL) model~\cite{Shigetani:1993dx}, the constituent quark model \cite{Frederico:1994dx,Kaur:2020vkq}, the chiral quark model \cite{Broniowski:2007si}, anti-de Sitter (AdS)/QCD models  \cite{Gutsche:2014zua,Gutsche:2013zia,Ahmady:2018muv,deTeramond:2018ecg}, a basis light-front quantization approach \cite{Lan:2019vui,Lan:2019rba} etc..

From the analyses of the Drell-Yan data, an essential feature of the pion valence PDF has been observed at high momentum fraction $x$. The pion valence PDF at large-$x$ shows a linear ${(1-x)^1}$ or slightly faster falloff, which is supported by the NJL model \cite{Shigetani:1993dx}, the constituent quark model \cite{Frederico:1994dx}, and duality arguments \cite{Melnitchouk:2002gh}. This observation is in discrepancy with perturbative QCD \cite{Farrar:1979aw,Berger:1979du,Brodsky:2006hj,Yuan:2003fs} and the Bethe-Salpeter equation approach \cite{Hecht:2000xa,Ding:2019lwe}, where the behavior has been predicted as ${(1-x)^2}$. The reanalysis of the data of the Drell-Yan process \cite{Aicher:2010cb}, including next-to-leading logarithmic threshold resummation effects, gives a considerably softer valence distribution at high-$x$  as compared to the NLO analysis.  

The Drell-Yan pair production in pion-nucleus scattering also provides information about the transverse structure of the pion in the momentum space. The differential cross section in the transverse momentum of the pair ($q_T$) for Drell-Yan pair production is a particularly suitable observable to get insight into the transverse structure of the pion. The description of the $q_T$ spectrum of the Drell-Yan pairs produced in pion-nucleus collisions has been recently reported in Ref. \cite{Wang:2017zym}. The analyses of the $q_T$ spectrum using different phenomenological models can be found in Refs. \cite{Pasquini:2014ppa,Ceccopieri:2018nop}.

Suitable wave functions for mesons are obtained in light-front holography, which relates the five dimensional AdS space to the Hamiltonian formulation of QCD on the light front \cite{stan2004, Brodsky2008a, teramond, stan2008, Brodsky:2008pf}. Although a rigorous QCD dual is unknown, a simple approach known as Bottom-Up allows to built the models that have some essential QCD features, such as counting rules and confinement. This formalism provides reasonable results for the meson sector \cite{ erlich, kruczen, rold, evans,witten, babin, sakai}. In this work, we first compute the pion valence quark PDFs by considering the light-front wave functions of two different models inspired by holographic QCD. We then evolve the initial valence quark PDFs of the pion within the next-to-next-to-leading order (NNLO) Dokshitzer-Gribov-Lipatov-Altarelli-Parisi (DGLAP) equations \cite{Dokshitzer:1977sg,Gribov:1972ri,Altarelli:1977zs} of QCD to the relevant scale in order to compare them with the PDF data from the E-0615 experiment at Fermilab. Further, using the pion PDFs in conjunction with the nuclear PDFs from the nuclear Coordinated Theoretical-Experimental Project on QCD (nCTEQ) 2015 global fit \cite{Kovarik:2015cma}, we investigate the cross section for the pion-nucleus induced Drell-Yan process in these models. We show that the PDFs from these models reasonably describe the measured data from a variety of experiments.

\section{Light-front holographic wave functions for the pion}\label{WF}
 The light-front wave functions (LFWFs) of the bound states are obtained as the eigenfunctions of the light-front Schr\"{o}dinger equation. In light-front holography, the Schr\"odinger equation for mesons is given by \cite{stan2008} 
\begin{equation}
\left(-\frac{\mathrm{d}^2}{\mathrm{d}\zeta^2}-\frac{1-4L^2}{4\zeta^2} + U_{\mathrm{eff}}(\zeta) \right) \phi(\zeta)=M^2 \phi(\zeta) \;,
\label{hSE}
\end{equation}
which is derived within a semiclassical approximation of light-front QCD. $\mathbf{\zeta}=\sqrt{x(1-x)} {b}_\perp$ is the holographic variable, where ${b}_\perp$ is the transverse distance between the quark and the antiquark, and $x$ is the light-front momentum fraction carried by the quark. $\mathbf{\zeta}$ maps onto the fifth dimension, $z$, of AdS space so that Eq. (\ref{hSE}) also describes the propagation of weakly-coupled string modes in a modified AdS space. The effective confining potential, $U_{\mathrm{eff}}$, is then obtained by the form of the dilaton field, $\varphi(z)$, which modifies the pure AdS geometry. Specifically, we have \cite{Brodsky:2014yha}
\begin{equation}
U_{\mathrm{eff}}(\zeta)= \frac{1}{2} \varphi^{\prime \prime}(z) + \frac{1}{4} \varphi^{\prime}(z)^2 + \frac{2J-3}{2 z} \varphi^{\prime}(z) \;,
\label{dilaton-potential}
\end{equation}	
where $J=L+S$. With the dilaton profile of the soft-wall model, $\varphi(z)=\kappa^2z^2$, one can solve the holographic Schr\"odinger equation, which yields the meson mass spectrum,
\begin{equation}\label{mass_spec}
M^2=4\kappa^2\left(n+{{J+L}\over2}\right),
\end{equation}
and the effective LFWFs,
\begin{equation}
\phi_{nL}(\zeta)=\kappa^{1+L}\sqrt{\frac{2 n !}{(n+L)!}}\zeta^{\frac{1}{2}+L}e^{\frac{-\kappa^2 \zeta^2}{2}}L_n^L(\kappa^2 \zeta^2),\label{phi-zeta}
\end{equation}
where $n$ and $L$ are the radial and the orbital quantum numbers. The first non-trivial prediction is that the lightest bound state, with quantum numbers $n=L=J=0$, is massless, i.e., $M^2=0$ from Eq.~(\ref{mass_spec}). Since spontaneous chiral symmetry breaking in massless QCD implies a massless pion, the ground state is naturally identified with the pion.

The holographic light-front Schr\"odinger equation only provides the transverse part of the meson LFWF. The complete wave function is given by \cite{Brodsky:2014yha}
\begin{equation}
\Psi (x,\zeta,\varphi)=\frac{\phi(\zeta)}{\sqrt{2\pi \zeta}} X(x) e^{iL\varphi} \;,
\end{equation}
where $X(x)=\sqrt{x(1-x)}$ is determined by a precise mapping of the electromagnetic form factor in AdS and physical space-time \cite{Brodsky:2008pf}. The normalized holographic LFWF for a ground state meson is then given by
\begin{equation}
\Psi (x,\zeta) = \frac{\kappa}{\sqrt{\pi}} \sqrt{x (1-x)}  e^{\frac{-\kappa^2 \zeta^2}{2}} \;.
\label{pionhwf} 
\end{equation}
Using $\zeta^2=x(1-x)\mathbf{b}_\perp^2$ in the Eq.~(\ref{pionhwf}) to replace $\zeta$, where the transverse impact variable $\mathbf{b}_\perp$ is conjugate to the light-front relative transverse momentum coordinate $\mathbf{k}_\perp$, the meson LFWF in momentum space can be written by performing a Fourier transform as
 \begin{equation}  \label{LFWF_kperp}
\psi(x,\textbf{k}_\perp)=\frac{4\pi N}{\kappa\sqrt{x(1-x)}} e^{-\frac{\textbf{k}_\perp^2}{2\kappa^2 x(1-x)} },
\end{equation}
where the normalization constant $N$ is fixed by
\begin{equation}
\int \frac{\mathrm{d}^2 \mathbf{k}_\perp \mathrm{d} x}{16\pi^3} ~|\psi(x,\textbf{k}_\perp)|^2 = 1 \;. 
\label{norm}
\end{equation}
This holographic model consists of massless quarks. A prescription was suggested in Ref. \cite{Brodsky2008a} to include quark masses, which was further developed into models to obtain meson wave function with massive quarks \cite{Vega2009, Vega2009a}.
\subsection{Model-I}\label{MI}
In Model-I, we account for non-vanishing quark masses in the meson LFWFs.
The quark and antiquark masses ($m_q$ and $m_{\bar{q}}$) are introduced by extending the kinetic energy of massless quarks or an equivalent change in Eq.~(\ref{hSE}),
\[
-\frac{d^2}{d\zeta^2}\rightarrow -\frac{d^2}{d\zeta^2} + \mu_{12}^2,
\]
which leads to
the mesons LFWF with the massive quarks \cite{Brodsky:2014yha,Swarnkar:2015osa},
\begin{equation}
\psi(x,{\bf k_{\perp}})=\frac{4\pi N}{\kappa_1 \sqrt{x(1-x)}} e^{-\frac{{\bf k}_{\perp}^2}{2 \kappa_1^2 x(1-x)}-\frac{\mu_{12}^2}{2 \kappa_1^2}},
\label{modelI}
\end{equation}
where $\mu_{12}^2=\frac{m_q^2}{x}+\frac{m_{\bar{q}}^2}{1-x}$. 
It has been observed that the light meson electromagnetic (EM) and transition ($\pi,~\eta,~\eta^{\prime}\to \gamma^*\gamma$) form factors with zero quark mass fail to agree with the experimental data. Meanwhile, the modified wave functions for massive quarks considerably improve the agreement with the data and the distribution amplitudes are also found to be consistent with pQCD predictions \cite{Swarnkar:2015osa}.

We use the modified holographic LFWF to evaluate the valence quark PDF of the pion. The probability of finding a quark inside the meson carrying the (longitudinal) momentum fraction $x$ is given by
\begin{equation}
f_{\rm I}(x)=\int ~ \frac{d^2{\bf k}_{\perp}}{16\pi^3}~|\psi(x,{\bf k_{\perp}})|^2,
\label{pdfI}
\end{equation}
which is interpreted as the PDF for the valence quark. The constituent quark mass and the scale parameter are only two parameters in this model.  We shall generate PDF in this work with the consistent quark mass i.e. $m_q=m_{\bar{q}}=330$ MeV, and the scale parameter $\kappa_1=540$ MeV, which lead to a good simultaneous description of a wide range of the pion data: the decay constant, charge radius, spacelike EM and transition form factors \cite{Swarnkar:2015osa}.
\subsection{Model-II}
\label{MII} 
Here, we consider the generalized form of the holographic LFWF of the pion by taking into account the quark orbital angular momentum. The LFWFs of pion, $  \psi^{L_{z}}_\pi(x,{\bf k}) $,  with total quark orbital angular momentum $L_{z} = 0,\pm1$ are given by \cite{Gutsche:2014zua}:
\begin{align}
\psi^{(0)}_{\pi}(x,{\bf k_{\perp}})=&\frac{4\pi N_0}{\kappa_2}\frac{\sqrt{\log(1/x)}}{1-x}\sqrt{f(x)\bar{f}(x)}\quad\quad \nonumber\\
&\times e^{-\frac{{{\bf k}_{\perp}^2}}{2\kappa_2^2}\frac{\log(1/x)}{(1-x)^2}\bar{f}(x)},\label{modelII1}\\
\psi^{(1)}_{\pi}(x,{\bf k_{\perp}})=&\frac{4\pi N_1}{\kappa_2}\frac{\sqrt{\log^3(1/x)}}{(1-x)^2}\sqrt{f(x)\bar{f}^3(x)}\nonumber \\
& \times e^{-\frac{{{\bf k}_{\perp}^2}}{2\kappa_2^2}\frac{\log(1/x)}{(1-x)^2}\bar{f}(x)},\label{modelII2}
\end{align}
where $ \kappa_2 $ is the scale parameter,  and $N_0$ and $N_1$ are the normalization factors. The original LFWF of the pion with $L_z=S_z=0$ \cite{meson-brodsky} has been extracted from light-front holography by matching the pion electromagnetic form factor in two approaches, AdS/QCD and light-front QCD. 
The LFWF adopted here is modified by introducing the profile functions in the original pion LFWF : 
\begin{align}
f(x)&=x^{\alpha-1}(1-x)^\beta(1+\gamma x^\delta), \nonumber\\
\bar{f}(x)&=x^{\bar{\alpha}}(1-x)^\beta(1+\bar{\gamma} x^{\bar{\delta}}),\label{profile}
\end{align}
 where $\alpha, \bar{\alpha}, \beta, \gamma, \bar{\gamma}, \delta, \bar{\delta}$ are the free parameters.
The profile function $f(x)$ is constrained by the pion valence quark distribution fixed through Ref.~\cite{Aicher:2010cb},
PDF $
\sim f(x)\,,  
$
while the other profile function $\bar f(x)$ is fixed from the analysis of 
the pion electromagnetic form factor.
The LFWFs in Eqs. (\ref{modelII1}) and (\ref{modelII2}) reduce to the original AdS/QCD LFWF when $f(x)= \bar{f}(x)=1$. The profile functions provide the correct scaling behavior of the pion PDF at large $x$ and the pion electromagnetic form factor at large momentum transfer. 

In particular, this parametrization of the pion wave functions yields the following scaling  of the PDF and form factor from the LFWF $\psi^{(0)}_\pi(x,\bf k)$:
$q(x) \sim (1-x)^2$  at $x \to 1$ and $F(Q^2) \sim 1/Q^2$ for $Q^2$ (momentum transfer) $\to \infty$. Meanwhile,
the LFWF $\psi^{(1)}_\pi(x,\bf k)$ results in 
$q(x) \sim (1-x)^5$
at $x \to 1$ and $F(Q^2) \sim 1/Q^4$ for $Q^2 \to \infty$. These are consistent with quark counting rules \cite{Gutsche:2014zua}.

We apply these parametrized wave functions in order to compute the pion valence quark PDF and to further study the cross section for the pion-nucleus induced Drell-Yan process in this model.  In term of the LFWFs, the pion valence quark PDF is expressed as 
\begin{align}
f_{\rm II}(x)=\int \frac{d^2{\bf k}_{\perp}}{16\pi^3}\Big[|\psi^{(0)}_{\pi}(x,{\bf k}_{\perp})|^2+{\bf k}_{\perp}^2|\psi^{(1)}_{\pi}(x,{\bf k}_{\perp})|^2\Big].
\label{pdfII}
\end{align}
The parameters $\alpha,~\beta,~\gamma$, and $\delta$ are taken from the Ref.~\cite{Aicher:2010cb}, whereas the parameters $\bar{\alpha},~\bar{\gamma}$, and $\bar{\delta}$ are fixed by a fit to data on the electromagnetic form factor of the pion with scale parameter $\kappa_2=350$ MeV \cite{Gutsche:2014zua}. With these parameters,
the modified LFWFs are able to reproduce several fundamental properties of the pion, such as the valence parton distribution, electromagnetic form factor, charge radius \cite{Gutsche:2014zua}, and GPDs \cite{Kaur:2018ewq}.   

Note that Model-I does not include the quark orbital angular momentum, while various pion observables such as the decay constant, charge radius, EM and transition form factors can be further improved by including the quark orbital angular momentum in this model~\cite{Ahmady:2018muv}. Furthermore, the transversely polarized quark TMD, also
known as the Boer-Mulders function, appears to be zero in Model-I without considering the quark orbital angular momentum~\cite{Ahmady:2019yvo}. 
The quark mass term in the exponential of Eq.~(\ref{modelI}) can be absorbed in the longitudinal mode for massive quarks. However, Model-II does not take into account the quark masses. Instead, the longitudinal modes have been modified by the profile functions given in Eq.~(\ref{profile})  which involve a set of parameters and successfully describe various properties of the pion. 
 It has been observed that Model-II works better than Model-I in order to describe the pion EM form factor~\cite{Gutsche:2014zua,Swarnkar:2015osa}. The pion form factor in Model-I is not consistent with quark counting rule, whereas  it is consistent with the rule in Model-II. The charge radius of the pion in Model-I and Model-II comes out to be $r^{\pi}_{\rm I}=0.529$ fm~\cite{Swarnkar:2015osa} and $r^{\pi}_{\rm II}=0.672$ fm~\cite{Gutsche:2014zua}, respectively, while the experimental value is $r^{\pi}_{\rm exp}=0.672\pm 0.008$ fm~\cite{Tanabashi:2018oca}.  
\begin{figure}[tbp]
\centering \includegraphics[width=.92\columnwidth]{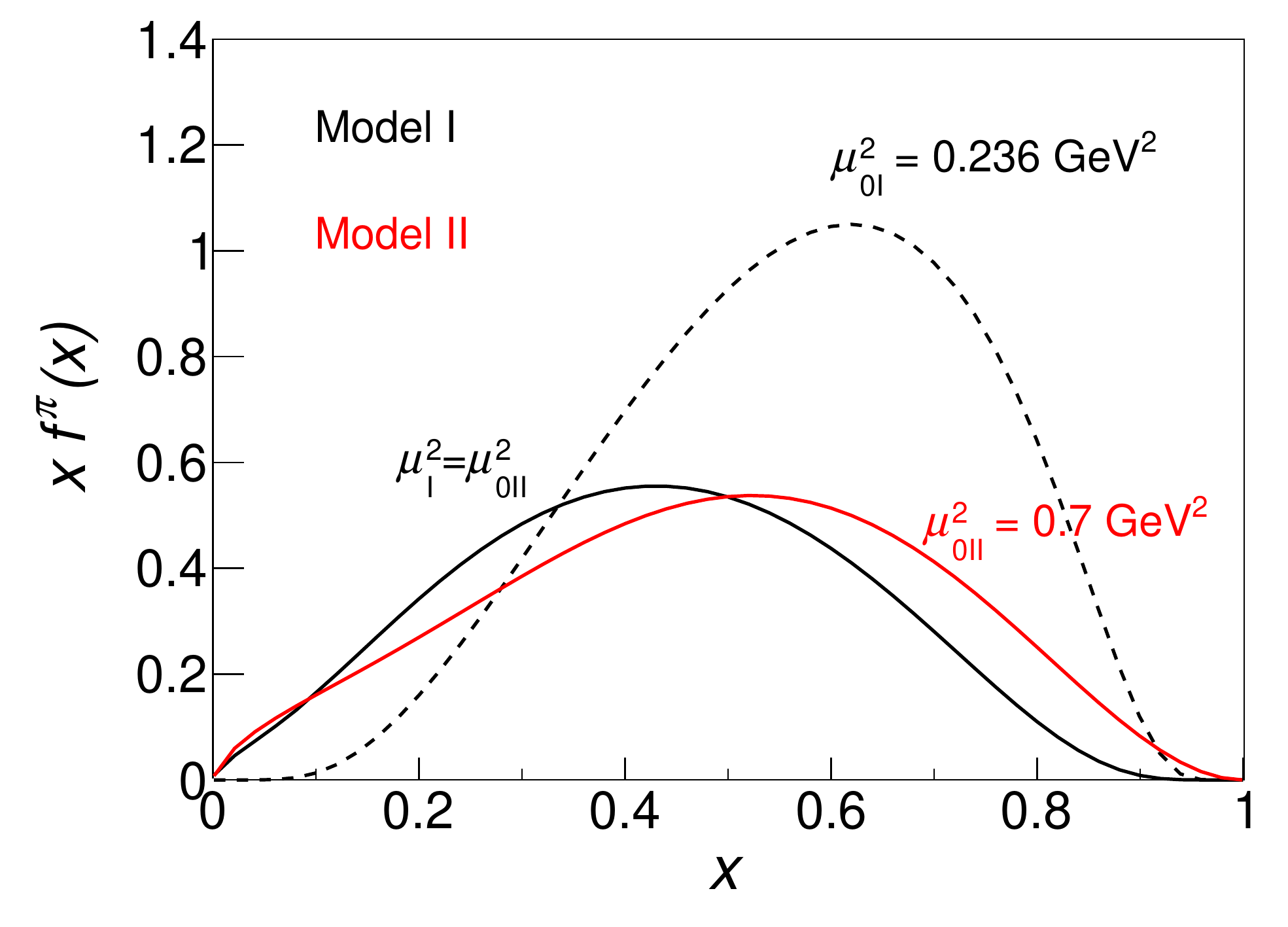}
\caption{Comparison of the valence quark PDFs in Model-I \cite{Brodsky:2014yha} and Model-II \cite{Gutsche:2014zua} at their model scales. The dashed black line represents the Model-I  at the model scale $\mu^2=0.236$ Gev$^2$. The solid black and the solid red lines correspond to the Model-I and the Model-II at the scale $\mu^2=0.7$ Gev$^2$, respectively.}
\label{initial}
\end{figure}
\subsection{QCD evolution of the PDFs}\label{dis_c}
The valence quark PDFs at a large scale can be obtained by performing the QCD evolution of the initial input PDF. We adopt the DGLAP equations \cite{Dokshitzer:1977sg,Gribov:1972ri,Altarelli:1977zs} 
of QCD with NNLO to evolve the PDFs from the initial scale, defined as $\mu_{0}^2$, to a larger scale $\mu^2$ as required for the comparison with experiment. Scale evolution allows (valence) quarks to emit gluons and the emitted gluons are further capable of producing the quark-antiquark pairs
as well as additional gluons. In other words, the larger scale reveals the gluon and the sea quark components of the constituent quarks through QCD interactions. 

Explicitly, we evolve the input PDFs given in Eqs. (\ref{pdfI}) and (\ref{pdfII}) to the relevant experimental scale $\mu^2=$ 16 GeV$^2$ using the higher order perturbative parton evolution toolkit (HOPPET) to solve the NNLO DGLAP equations~\cite{Salam:2008qg}. While applying the DGLAP equations numerically, we impose the condition that the running coupling $\alpha_s(\mu^2)$ saturates in the infrared at a cutoff value of max $\alpha_{s}=1$ \cite{Lan:2019vui,Xu:2019xhk}. We adopt the initial scale ${\mu_{0\rm I}^2=0.236\pm0.024~\rm{GeV}^2}$ for the Model-I, which we determine by requiring the results after QCD evolution to fit the reanalysis of the FNAL-E-0615 data for the pion PDF~\cite{Aicher:2010cb,Chen:2016sno}. Similarly, for the Model-II, the initial scale is obtained as $\mu_{0\rm II}^2=0.70\pm0.07$ GeV$^2$.  At the central value of the initial scales, the $\chi^2$ per degree of freedom for the fit of the pion PDF in the Model-I is $3.12$, whereas in the Model-II, the value is $3.62$. We estimate a $10\%$ uncertainty in the initial scales.

\begin{figure}[tbp]
\centering \includegraphics[width=.92\columnwidth]{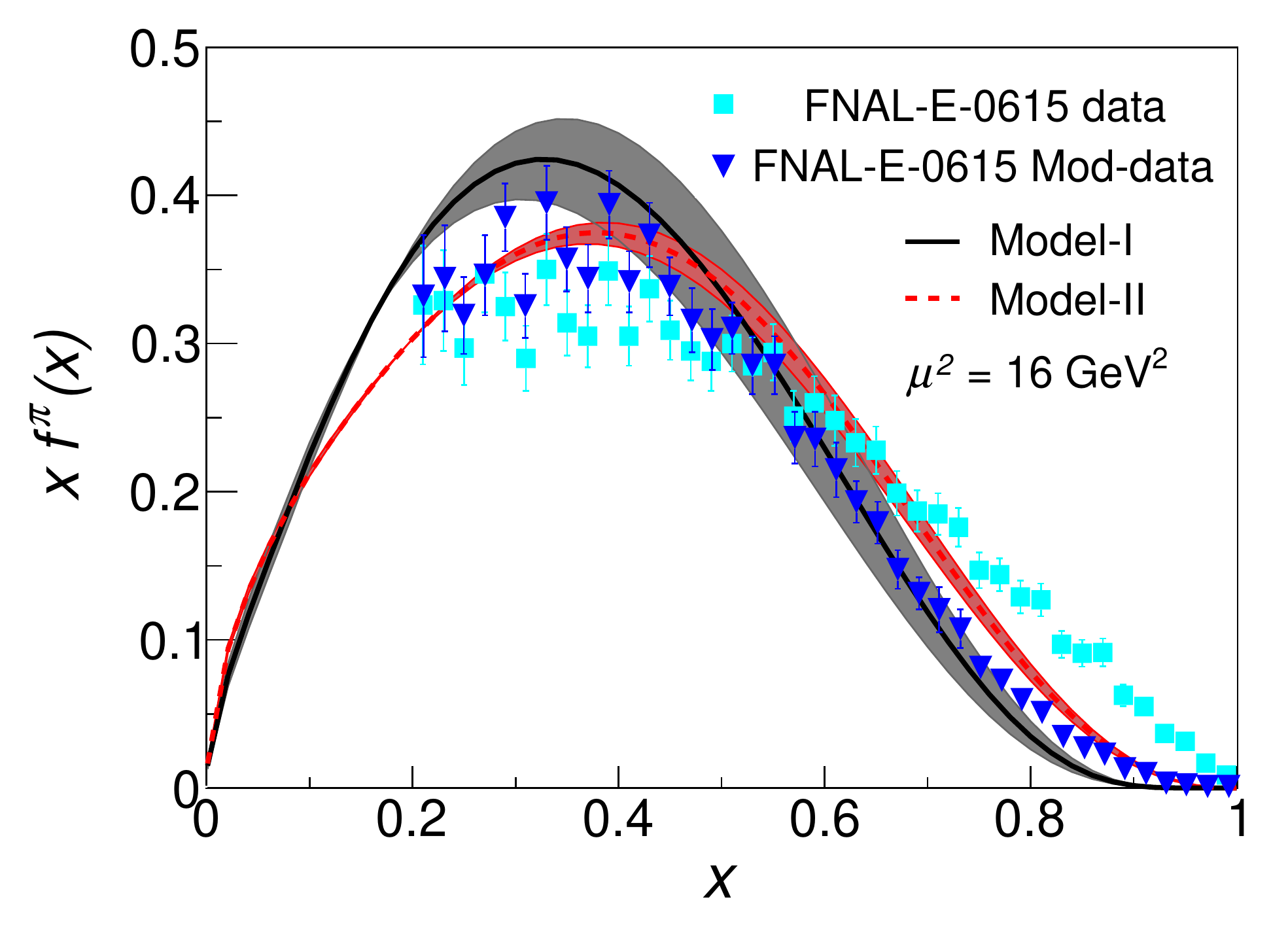}
\caption{Plot of $xf^\pi(x)$ for valence quark as a function of $x$. The solid black and dashed red lines correspond to the Model-I \cite{Brodsky:2014yha} and the Model-II \cite{Gutsche:2014zua}, respectively. The FNAL-E-0615 experiment data are taken from \cite{Conway:1989fs}. The FNAL-E-0615 modified data are taken from \cite{Aicher:2010cb,Chen:2016sno}.
}
\label{fpionE615}
\end{figure}
 We show the valence quark PDFs in Model-I and Model-II at their initial scales in Fig.~\ref{initial}. We find that the valence quark and antiquark in Model-I carry the entire light-front momentum of the pion, which is appropriate for a low-resolution model. However, in Model-II, the valence quark and antiquark together carry $\sim$ $62\%$ of the pion momentum, which justifies to fix the larger scale, $\mu_{0\rm II}^2 = 0.7$ GeV$^2$, compared to Model-I. In order to compare Model-I with Model-II, we evolve the PDF in Model-I from its initial scale to $\mu_{0\rm II}^2 = 0.7$ GeV$^2$. We notice that at this scale, the valence quark and antiquark together carry $\sim$ $60\%$ of the pion momentum in Model-I. Therefore, a compatible agreement has been observed between the PDFs in these two models.

In Fig. \ref{fpionE615}, we show the valence quark PDFs after QCD evolution and compare them with the data from the E-0615 experiment at Fermilab for the Drell-Yan process \cite{Conway:1989fs}  as well as with the reanalysis of the data which includes next-to-leading logarithmic threshold resummation effects \cite{Aicher:2010cb,Chen:2016sno}. Note that at large momentum fraction $x$, the data for the Drell-Yan process shows typically a linear $(1-x)^1$ or slightly faster falloff. However, the reanalysis of the data exhibits the pion PDF as $(1-x)^2$ at large-$x$. This behavior is in agreement with the prediction from the calculations using Dyson-Schwinger equations \cite{Hecht:2000xa}. Here, we observe that the pion valence PDF in Model-I falls off as $(1-x)^{2.33}$, while the PDF in Model-II falls off as $(1-x)^{1.65}$. Overall, the Model-II shows better agreement than the Model-I when we compare them with both the data. Again, Model-I describes the modified data better than Model-II within the range $0.5<x<0.8$.  

\begin{figure}[tbp]
\centering \includegraphics[width=.9\columnwidth]{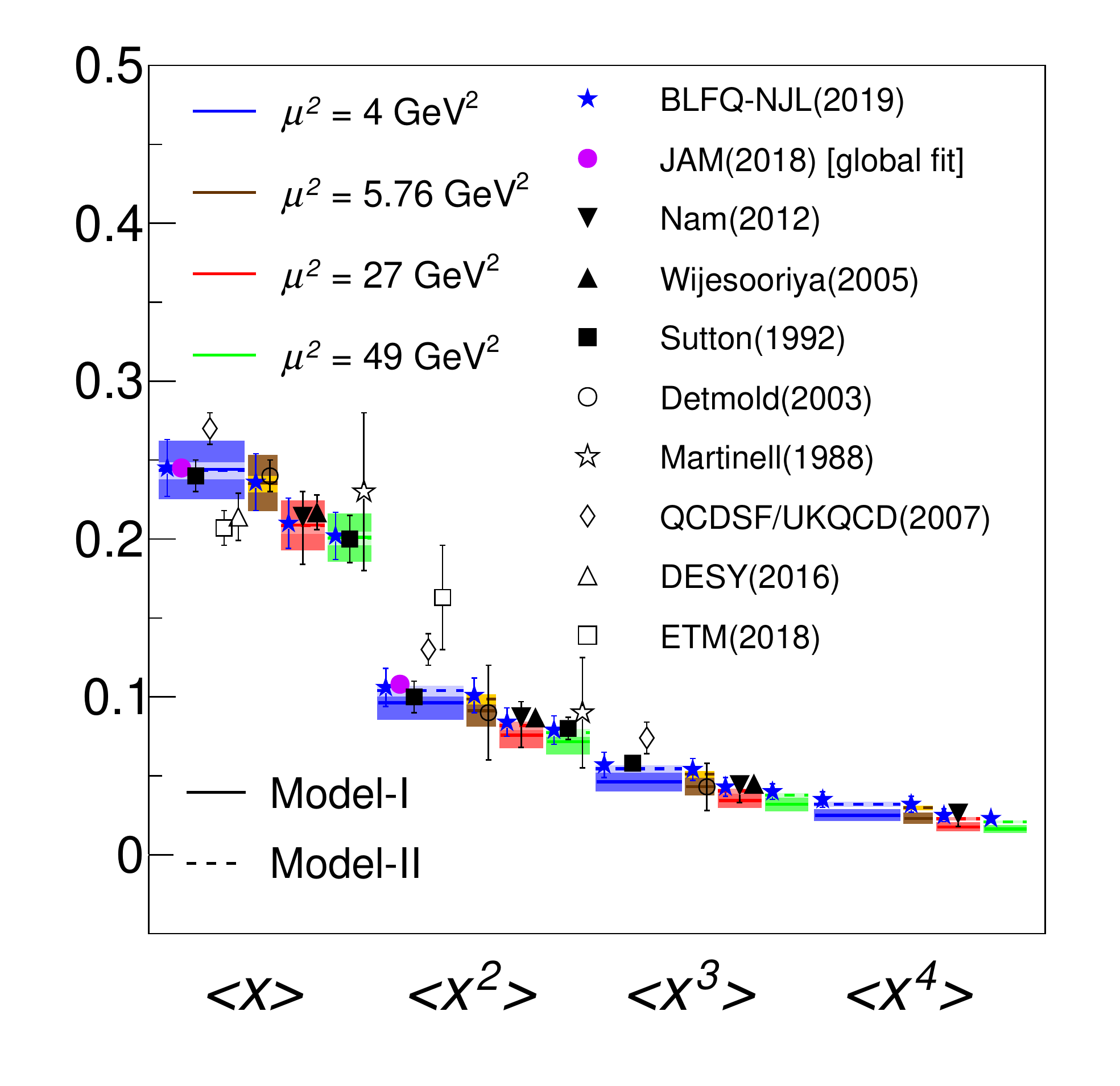}
\caption{Comparison of the lowest four moments of valence quark PDF in the pion with the global fit \cite{Barry:2018ort}, with lattice QCD results in Refs.~\cite{Detmold:2003tm,Martinelli:1987bh,Oehm:2018jvm,Abdel-Rehim:2015owa,Brommel:2006zz}, and with phenomenological models in Refs.~\cite{Sutton:1991ay,Wijesooriya:2005ir,Nam:2012vm,Lan:2019vui,Lan:2019rba} at different scales. The colored horizontal bars are results of the present work. Solid and dashed lines represent the Model-I and Model-II, respectively. The bands with deep and light color correspond to the error bands in Model-I and Model-II, respectively.}
\label{moment}
\end{figure}

To compare with the other theoretical results, we evaluate the lowest four nontrivial moments of the valence quark PDF :
\begin{equation}
\langle x^{n} \rangle=\int_0^1 dx~ x^n f_{\rm v}^{\pi}(x,\mu^2),~n=1,2,3,4.
\label{pionmoment1}
\end{equation}
In Fig. \ref{moment}, we show the first four moments of valence quark PDF of the pion at different $\mu^2$ and compare them with the results from the global fit \cite{Barry:2018ort}, lattice QCD results in Refs. \cite{Detmold:2003tm,Martinelli:1987bh,Oehm:2018jvm,Abdel-Rehim:2015owa,Brommel:2006zz}, the BLFQ-NJL model \cite{Lan:2019vui,Lan:2019rba}, and other phenomenological models in Refs. \cite{Sutton:1991ay,Wijesooriya:2005ir,Nam:2012vm}. We observe that both the models agree  with the theoretical results in Refs. \cite{Sutton:1991ay,Detmold:2003tm,Wijesooriya:2005ir,Nam:2012vm,Barry:2018ort,Martinelli:1987bh,Lan:2019vui,Lan:2019rba}. However, the Model-II shows a better agreement compared to the Model-I for the higher moments. One can notice that the moments of the valence quark decrease as the scale $\mu^2$ increases, which implies increasing contributions from sea quarks and gluons at larger scales.

\begin{figure*}[tbp]
\centering (a)\includegraphics[width=.92\columnwidth]{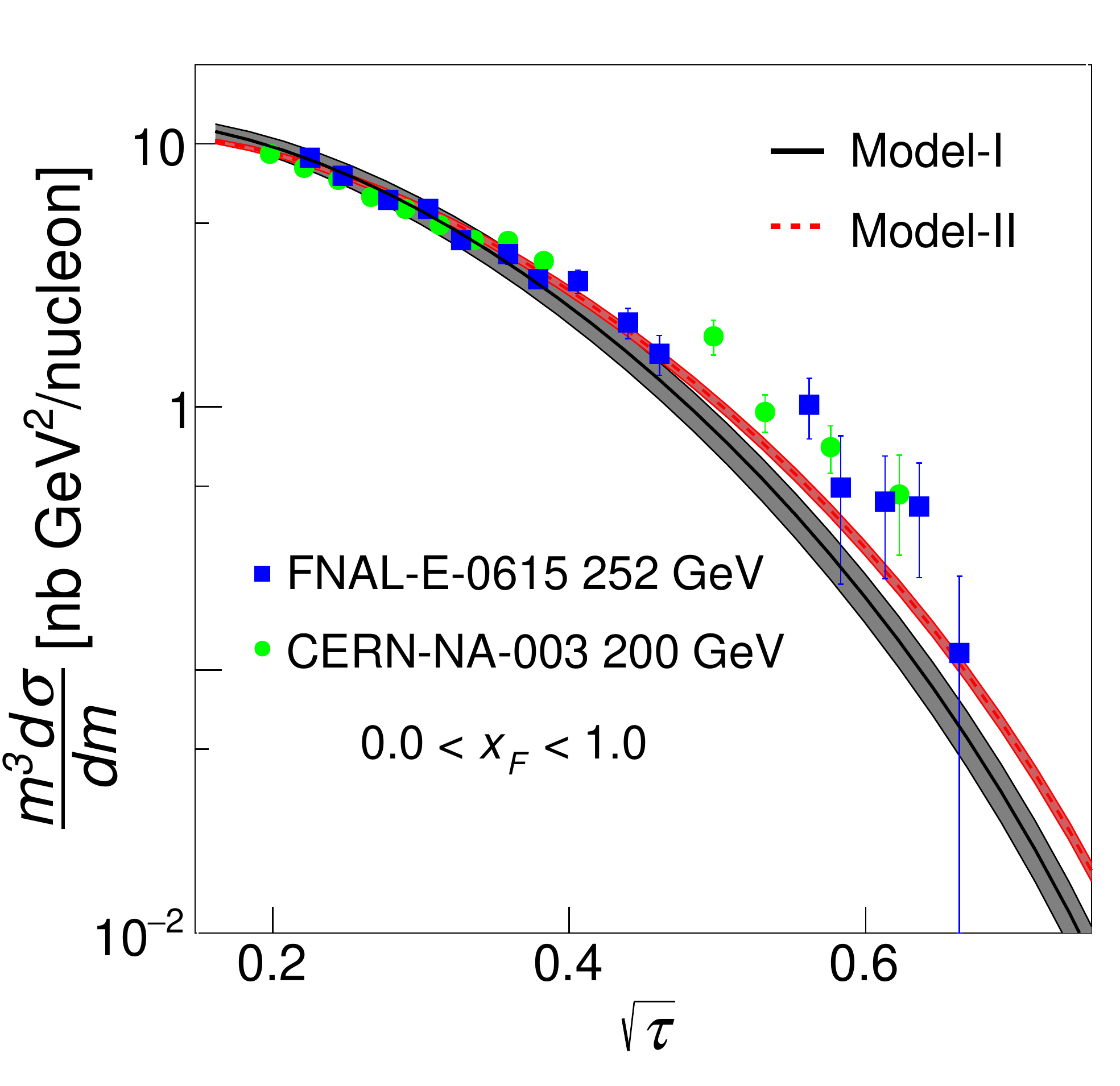}
(b)\includegraphics[width=.92\columnwidth]{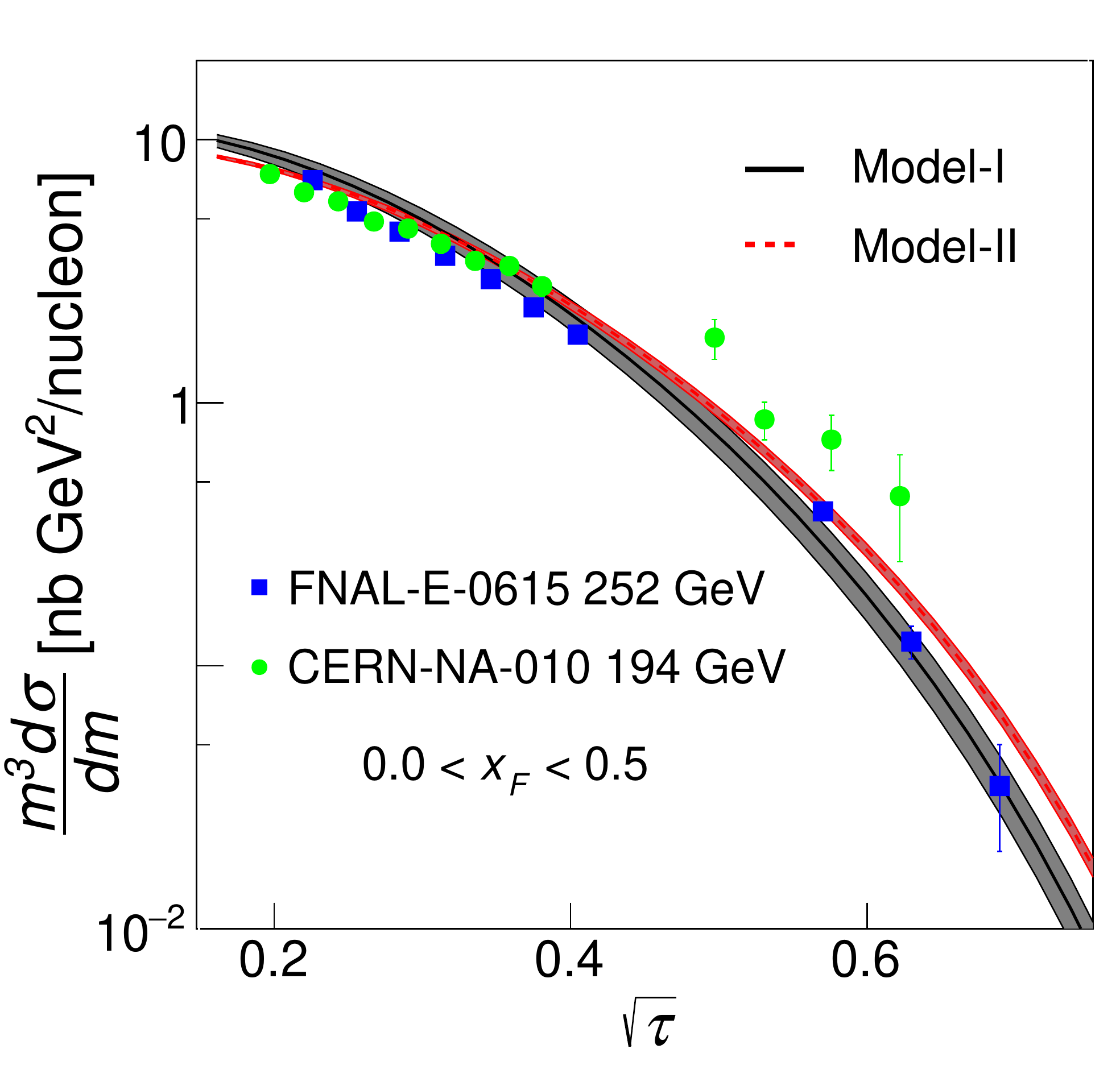}
\caption{Plot of $m^3d{\bf \sigma}/dm$ as a function of $\sqrt{\tau}$ for the region (a) $0<x_{\rm F}< 1$ and (b) $0< x_{\rm F}< 0.5$. The data of the FNAL-E-0615 experiment (tungsten target) with 252 GeV pions, the CERN-NA-003 experiment (platinum target) with 200 GeV pions, and the CERN-NA-010 experiment (tungsten target) with 194 GeV pions are taken from Ref. \cite{Conway:1989fs}, Ref. \cite{Badier:1983mj}, and Ref. \cite{Betev:1985pf}, respectively.}
\label{cross1}
\end{figure*}

\section{Cross section in unpolarized Drell-Yan process}
\label{SecIII}
Here, we present the cross section for the Drell-Yan process~\cite{Drell:1970wh} using the pion PDFs evaluated in those two models. 
We define the momenta of the incoming hadrons and the outgoing lepton pair as $p_{1,2}$ and $l_{1,2}$, respectively. 
The kinematics of the process are described by the center of mass energy square of incoming hadrons $s$, the invariant mass of the lepton pair $m$, rapidity $Y$, the Feynman variable $x_{\rm F}$, and the variables $\tau$, $z$ and $y$. The variables are related to each other and defined as \cite{Becher:2007ty}:
\begin{align}
&s = (p_1+p_2)^2,  \quad\quad q=l_1+l_2,	\nonumber\\
&m^2 =  q^2,\quad\quad Y=\frac12\,\ln\frac{q^0+q^3}{q^0-q^3},\nonumber\\
&   x_{\rm F} = x_1 - x_2,\quad\quad	\tau \equiv \frac{m^2}{s},\\
&   z = \frac{m^2}{\hat{s}}=\frac{\tau}{x_1x_2},\quad\quad	y=\frac{\frac{x_1}{x_2}e^{-2Y}-z}{(1-z)(1+\frac{x_1}{x_2}e^{-2Y})}, \nonumber
\end{align}
 where, $\hat{s}=x_1x_2s$. $x_i$ represent the fractions of the hadron momenta $p_i$ carried by the annihilating quark or antiquark, and are given by in term of other variables,
 \begin{eqnarray}
\label{Eq:x1-x2}
 &&     x_{1}= \sqrt{\frac{\tau}{z}\frac{1-(1-y)(1-z)}{1-y(1-z)}}e^{Y}, \nonumber\\
&&      x_{2}= \sqrt{\frac{\tau}{z}\frac{1-y(1-z)}{1-(1-y)(1-z)}}e^{-Y}.
\end{eqnarray}

The cross section for the Drell-Yan process: $\pi^-$ $\rm{Nucleus}$ $\rightarrow \mu^+\mu^-X$ can be determined in perturbative QCD and expressed in terms of convolutions of short-distance partonic cross sections with PDFs as
\cite{Becher:2007ty,Anastasiou:2003yy,Anastasiou:2003ds,Barry:2018ort},
\begin{align}
&\frac{m^3 d^2{\bf \sigma}}{dm\, dY}=\frac{8\pi\alpha^2}{9 }\frac{m^2}{s}\sum_{ij} \int dx_1 dx_2 \nonumber\\ &\times \widetilde{C}_{ij}(x_1,x_2,s,m,\mu^2) f_{i/\pi}(x_1,\mu^2) f_{j/N}(x_2,\mu^2),%
\label{crosseq}
\end{align}
where $\widetilde{C}_{ij}$ are the hard-scattering kernels, which have an expansion in powers of the strong coupling $\alpha_s$. The sums go over all possible partonic channels contributing at a given order in the expansion of $\widetilde{C}_{ij}$. At leading order ($\sim\alpha_s^0$), only the channels $(ij)=(q\bar q), (\bar q q)$ contribute, whereas at NLO ($\sim\alpha_s$), one must include $(ij)=(q\bar q), (\bar q q), (qg), (gq), (\bar q g), (g\bar q)$ in the sum. The hard-scattering kernels at NLO can be found in Ref. \cite{Becher:2007ty}. In this work, we present the cross section up to NLO. In order to evaluate the cross sections in Eq. (\ref{crosseq}), in conjunction with the pion PDFs at the experimental scale $\mu^2=16$ GeV$^2$, we adopt the nuclear PDFs from the nCTEQ 2015 \cite{Kovarik:2015cma} at the same scale. The PDFs for the tungsten and the beryllium nuclei are readily available in Ref. \cite{Kovarik:2015cma}, while for the PDFs in the platinum nucleus, we approximate them by the corresponding bound nucleon PDFs in the gold nucleus.

\begin{figure*}[tbp]
\centering (a)\includegraphics[width=.92\columnwidth]{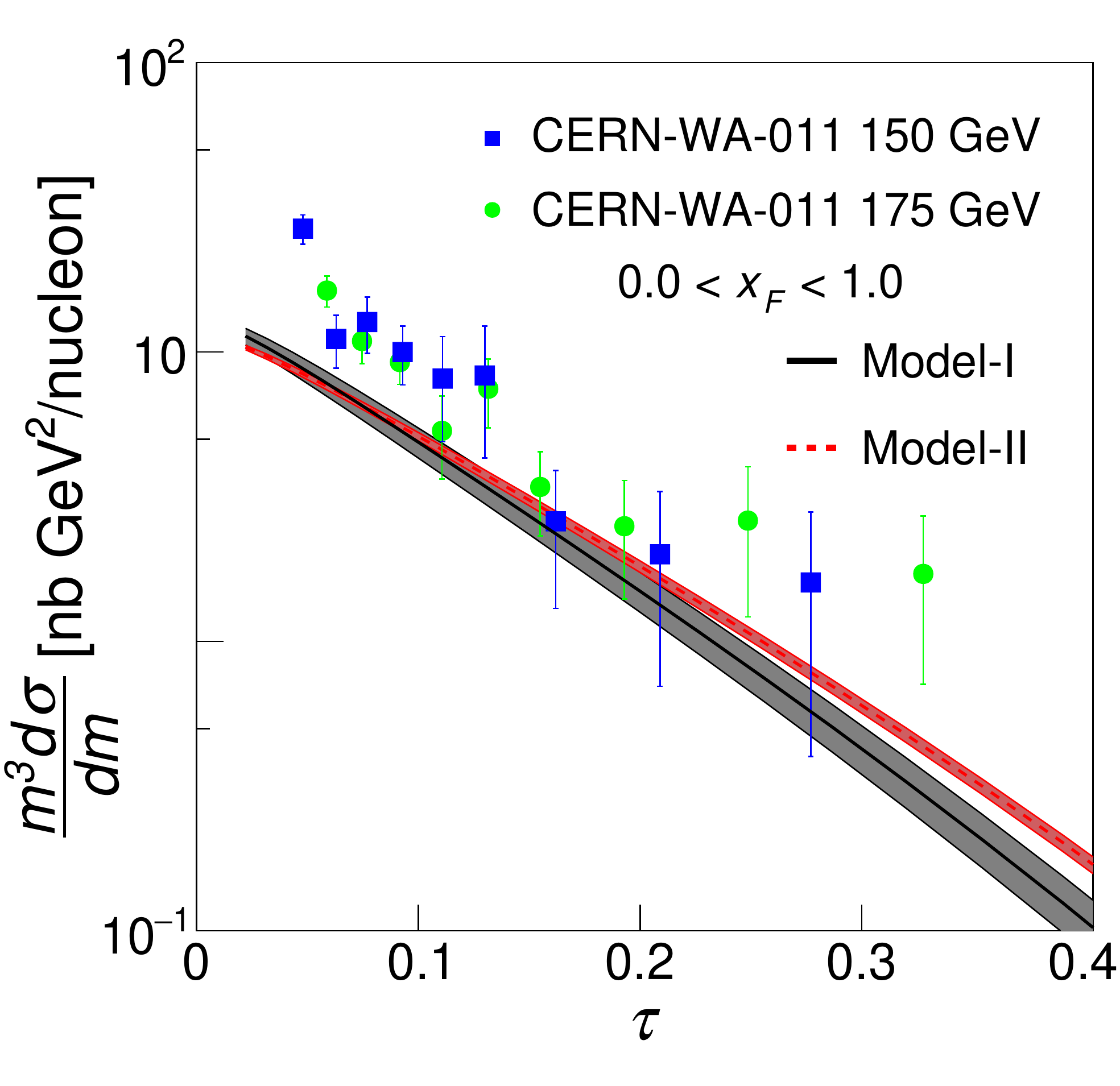}
(b)\includegraphics[width=.92\columnwidth]{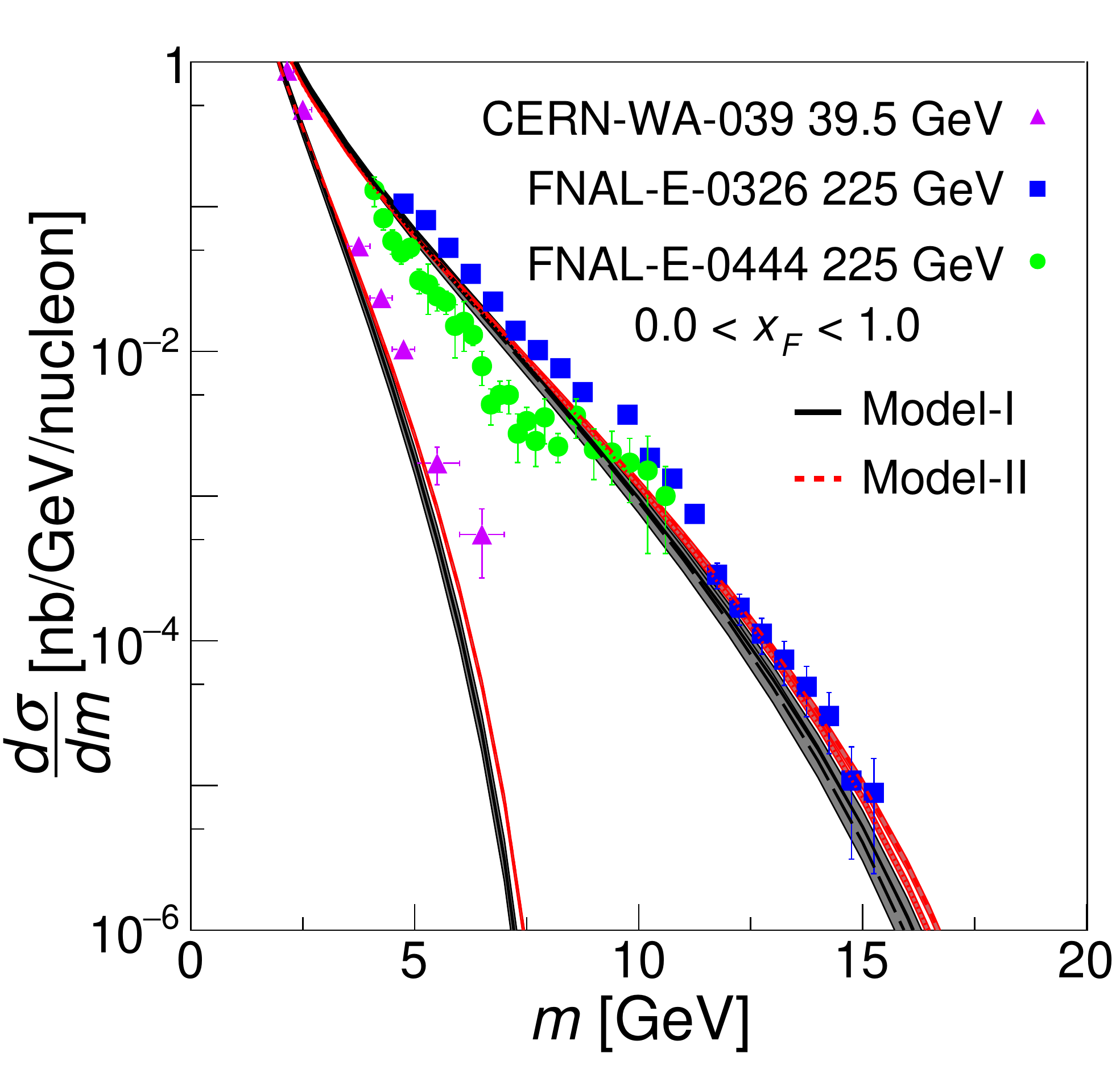}
\caption{Plot of (a) $m^3 d{\bf \sigma}/dm$ as a function of $\tau$ for the region  $0<x_{\rm F}< 1$. The data of the CERN-WA-011 experiment (beryllium target) with 150 GeV and 175 GeV pions are taken from Ref. \cite{Barate:1979da}; (b) $ d{\bf \sigma}/dm$ as a function of $m$ for the same region. The data of the FNAL-E-0326 experiment (tungsten target) with 225 GeV pions and the FNAL-E-0444 experiment (carbon target) with 225 GeV pions are taken from Ref. \cite{Greenlee:1985gd} and Ref. \cite{Anderson:1979tt}, respectively. The data of the CERN-WA-039 experiment (tungsten target) with 39.5 GeV pions are taken from Ref. \cite{Corden:1980xf}.}
\label{cross3}
\end{figure*}

We show the differential cross sections after integrating out the $Y$ dependence of the ${m^3 d^2{\bf \sigma}/dm \,dY}$ in Fig.~\ref{cross1} and Fig.~\ref{cross3}(a) as a function of $\sqrt{\tau}$ and $\tau$, respectively, while in Fig.~\ref{cross3}(b), we present the cross section, ${d{\bf \sigma}/dm }$, as a function of $m$ to compare them with the experimental data. In Fig.~\ref{cross1}(a), the FNAL-E-0615 and the CERN-NA-003 data correspond to the tungsten and the platinum targets, respectively. The data in Fig.~\ref{cross1}(b) correspond to the tungsten target. However, we employ the tungsten nuclear PDFs to compute the cross section in Fig.~\ref{cross1}. We notice that per nucleon cross section evaluated by considering the tungsten and the platinum nuclear PDFs are very close. Since the CERN-WA-011 data in Fig.~\ref{cross3}(a) represent the beryllium target~\cite{Barate:1979da}, we use the same target nuclear PDFs to evaluate the cross section shown in this plot. The cross section $d{\bf \sigma}/dm $ as a function of $m$ displayed in Fig.~\ref{cross3}(b) are compared with the data of the CERN-WA-039 experiment with 39.5 GeV pions~\cite{Corden:1980xf} and with the data of the FNAL-E-0326 experiment~\cite{Greenlee:1985gd} as well as the FNAL-E-0444 experiment~\cite{Anderson:1979tt} with 225 GeV pions. The FNAL-E-0326 and the CERN-WA-039 data correspond to the tungsten target, whereas the FNAL-E-0444 data represent the carbon target. The corresponding targets of nuclear PDFs are used to obtain the cross section shown in  Fig. \ref{cross3}(b). The Figs. \ref{cross1} and \ref{cross3} suggest that both the holographic inspired models show a reasonable agreement with data from widely different experimental conditions.

\section{Summary}\label{SecIV}
We calculated the valence quark PDF of the pion using the pion light-front wave functions in two different models inspired by light-front holographic QCD.  The initial scales of the PDFs have been obtained by consistently fitting the evolved pion valence quark PDFs to the modified E-0615 data~\cite{Aicher:2010cb,Chen:2016sno}. Our analysis shows that the Model-II is better than Model-I when compared with both the E-0615 data and the reanalyzed data by including next-to-leading logarithmic threshold resummation effects. The moments of the pion PDFs have been found in consistent with the global fit by JAM Collaboration~\cite{Barry:2018ort}, with lattice QCD~\cite{Detmold:2003tm,Martinelli:1987bh}, as well as with phenomenological quark models~\cite{Sutton:1991ay,Wijesooriya:2005ir,Nam:2012vm,Lan:2019vui,Lan:2019rba} across various scales. However, we found that the Model-II shows a slightly better agreement compared to Model-I for higher moments.

We have investigated the cross sections up to NLO of the pion-nucleus induced Drell-Yan process using the pion PDFs of these two models.
In conjunction with the pion PDFs, we employed the nuclear PDFs from the nCTEQ 2015~\cite{Kovarik:2015cma} to evaluate the cross sections.  In comparison with the data from widely different experimental conditions ~\cite{Conway:1989fs,Badier:1983mj,Barate:1979da,Corden:1980xf,Greenlee:1985gd,Anderson:1979tt}, we observed that both models provide an acceptable description of various experimental data.

\section{ Acknowledgements}
We thank Xingbo Zhao, Shaoyang Jia, Ruben Sandapen and Mohammad Ahmady for many useful discussions. We   also thank Satvir Kaur for critically
reading the manuscript and giving valuable suggestions. This work is supported by the National Natural Science Foundation of China (NSFC) under the Grants No. 11850410436 and No. 11950410753. C.M. is supported by new faculty startup funding by the Institute of Modern Physics, Chinese Academy of Sciences. This work is also supported by the Strategic Priority Research Program of Chinese Academy of Sciences, Grant No. XDB34000000 and Key Research Program of Frontier Sciences, CAS, Grant
No ZDBS-LY-7020.

\end{document}